\documentclass[%
reprint,
showpacs,
 amsmath,amssymb,
 aps,pra,
]{revtex4-1}

\usepackage{graphicx}
\usepackage{dcolumn}
\usepackage{bm}

\usepackage{gensymb}
\usepackage{subfigure}
\usepackage{flushend}

\begin{document}

\title{Resonance Retrieval of Stored Coherence
 in a Radiofrequency-optical Double Resonance Experiment }

\author{Vladimir Djokic}
\author{Georg Enzian}%
\author{Frank Vewinger}
\author{Martin Weitz}
\affiliation{Institut f{\"u}r Angewandte Physik der Universit{\"a}t Bonn, Wegelerstr. 8, D-53115 Bonn, Germany
}%

\begin{abstract}
We study the storage of coherences in atomic rubidium vapor with a three-level coupling scheme with two ground states and one electronically excited state driven by one optical (control) and one radiofrequency field. We initially store an atomic ground state coherence in the system. When retrieving the atomic coherence with a subsequent optical pulse, a second (signal) optical beam is created whose difference frequency to the control field is found to be determined by the atomic ground states Raman transition frequency.
\end{abstract}

\maketitle



In the past decades quantum coherence effects have been a topic of intense scientific research. 
Light-matter interactions that induce coherent effects have a remarkable capability of being able to alter optical properties of atomic media. Of crucial importance in many fields, e.g. in quantum information, is the ability to coherently manipulate atomic states and retrieve previously mapped photon states. Electromagnetically induced transparency (EIT) \cite{Fleisch2005} and coherent population trapping \cite{Alz} are widely used tools that allow for reversible storage of light in atomic ensembles \cite{Liu,Lukin1}, where stored photonic information can be controlled in a coherent manner \cite{Lukin2}. Based on the EIT effect, it has been reported that non-classical states of light, like squeezed and single-photon states, have been reversibly stored in atomic ensembles \cite{Appel,Eisman,SqueezedVacuum}. Application of EIT in quantum devices and generation of photonic qubits has also been proposed \cite{qubit,Frank}. Besides the EIT phenomenon, many different storage techniques have been proposed, such as off-resonant Raman interaction \cite{offreson} or gradient echo memories (GEM) \cite{GEM} schemes in both gaseous atomic samples \cite{gematom1,gematom2} and ion-doped crystals \cite{Nilsson}. The possibility to prepare narrow resonances using EIT \cite{narrow1,narrow2} makes it a valuable tool for magnetometry \cite{mag1,mag2,mag3} and optical clocks. State-of-the-art magnetometry that surpasses the precision of superconducting quantum interference devices has been reported \cite{mag4}. Of furhter interest is the application of coherent population trapping in the measurement of the frequency difference between two hyperfine atomic states allowing for miniature atomic clocks \cite{clock1,clock2}. While many experiments use continuous signal read out, other work has used free spin precessing for atomic magnetometry, resulting in a Ramsey-like scheme \cite{weis1}.

In the present manuscript, we examine the eigenfrequency of the retrieved atomic coherence following initial storage of coherence, inspired by light storage experiments in dark state media \cite{Lukin1}. In previous works using light storage we have demonstrated that the difference frequency of the regenerated signal beam to the control beam used for storage oscillated at an atomic eigenfrequency \cite{leon, leon1}. It is well known that atomic ground state coherences can be created also in an optical-radiofrequency double resonance configuration \cite{quantumbeat}. Here, we demonstrate resonance retrieval of atomic coherence in an optical-radiofrequency double resonance experiment. Our experiment uses an atomic three-level system, and the setup is based on hot atomic rubidium vapor used for the storage. After the retrieval of the stored coherence, a beating that matches the energy difference between the two ground levels is observed. Within our experimental uncertainty, the observed beat frequency is insensitive to a wide range of variations of both the drive frequency $\omega_\text{rf}$ and the amplitude of the radiofrequency (rf) field. The observed effect is robust with respect to variations of the Rabi frequencies of both optical control and rf fields. 
\begin{figure}[b!]
\centering
\includegraphics{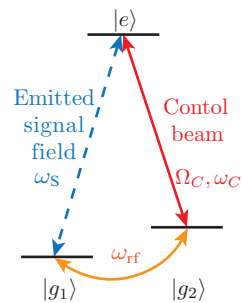}
\caption{\label{fig:3sys} Simplified three-level scheme used
for resonant retrieval of stored coherence. The levels  $| g_2\rangle$
and  $|e\rangle$ are coupled by a control field with the Rabi frequency 
$\Omega_C$ while a rf field at frequency $\omega_\text{rf}$ couples the ground levels 
$|g_1\rangle$ and $|g_2\rangle$. After storage of the ground state coherence, an optical signal field can be generated with the difference of the optical fields oscillating at the (Raman) transition frequency between ground states $|g_1\rangle$ and  $|g_2\rangle$.}
\end{figure}

For a simple model of our experimental method, consider an ensemble of three-level atoms as depicted in Fig.~\ref{fig:3sys} with two stable ground states $|g_1\rangle$, $|g_2\rangle$ and one electronically excited state $|e\rangle$ that is spontaneously decaying. The transition between the states $|g_2\rangle$ and $|e\rangle$ is coupled coherently via the optical control laser beam with Rabi frequency $\Omega_C$, while the rf field at frequency $\omega_\text{rf}$ couples the ground levels $|g_1\rangle$ and $|g_2\rangle$. Without the presence of the rf coupling, the system is pumped into the ground state $|g_1\rangle$. Adding of the rf field coupling creates a coherence among ground levels and together with the simultaneous driving of the optical transition induces coherence between levels $|g_1\rangle$ and $|e\rangle$, that in turn produces a coherent optical signal field collinear with the control laser beam. During the rf excitation of the system, the generated signal field beats with the transmitted laser field at the frequency $\omega_\text{beat}=\omega_S-\omega_C =\omega_\text{rf}$.
We show that after the simultaneous switching off of the optical and rf excitation fields, and the later resonant retrieval of the stored coherence, the retrieved light beats at the difference frequency of the ground state levels $\omega_\text{beat}=\omega_S-\omega_C =\omega_{g_1e}-\omega_{g_2e}$.

\begin{figure}[b]
\includegraphics{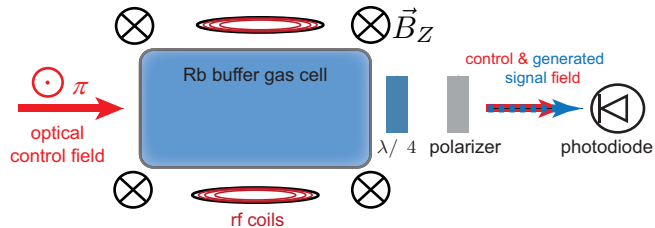}
\caption{\label{fig:setup}Scheme of the experimental setup.}
\end{figure}

\begin{figure}[t!]
\includegraphics{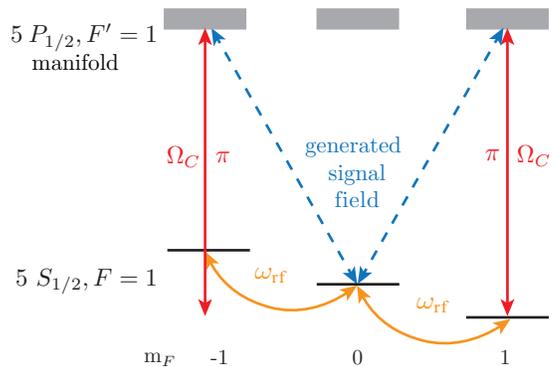}
\caption{\label{fig:6sys}Full level scheme of the experimentally investigated $\text{Rb}^{87}$  $F=1 \rightarrow F'=1$ transition, as also used in our numerical calculations. The solid blue and orange lines represent the transitions driven by the $\pi$-polarized optical control field and the rf field respectively. The dashed blue lines represent the coupling of the generated $\sigma^+ - \sigma^-$ polarized optical signal field.}
\end{figure}

In our experimental apparatus (Fig.~\ref{fig:setup}) a grating stabilized diode laser is used as a source of the optical control beam. The diode laser is locked with dichroic atomic vapor laser lock procedure \cite{davll} to the ${5 S_{1/2}\ F=1 \rightarrow 5 P_{1/2}\ F'=1}$ (see Fig.~\ref{fig:6sys}) hyperfine component of the rubidium D1 line near 795 nm. Its emission passes an acousto-optic modulator, is spatially filtered with a fiber, and then with linear polarization sent to a magnetically shielded 50 mm long rubidium buffer gas cell. The 
cell itself contains 1 Torr xenon buffer gas and is heated up to $80\degree$C, resulting in a rubidium atom number density of $\approx 10^{-12} \text{cm}^{-3}$. To lift the degeneracy of the Zeeman sublevels of the $F=1$ ground state, a magnetic bias field is applied in a direction parallel to the control beam polarization.

The splitting between the adjacent Zeeman sublevels of the ground state is  $g_F\mu_BB$, where $\mu_B$ is the Bohr magneton, and the hyperfine $g_F$ factor equals 1/2. In our experimentally used level scheme, the ground state sublevels $|F = 1, m_F = 0\rangle$ and $|F = 1, m_F =\pm1\rangle$ correspond to the states $|g_1\rangle$ and $|g_2\rangle$ of the  three-level model presented in the Fig.~\ref{fig:3sys} respectively,  and $|F' = 1, m_F = \pm1\rangle$ of the electromagnetically excited state to $|e\rangle$. The $\pi$-polarized optical field drives the transition $|F = 1, m_F =\pm1\rangle$ to $|F' = 1, m_F =\pm1\rangle$, whereas the generated signal beam is $\sigma_+-\sigma_-$ polarized and tuned near the $|F' = 1, m_F =\pm1\rangle$ to $|F = 1, m_F = 0\rangle$ transition. The rf field coherently couples the Zeeman states of the $5 S_{1/2}, F=1$ ground state manifold. This field is generated by a radio-frequency antenna mounted close to the rubidium cell. In the absence of a radiofrequency field, the linearly polarized control field pumps the atomic population into the $|F = 1, m_F = 0\rangle$ ground state sublevel. After leaving the rubidium cell, the control field and the generated signal field are projected onto the same polarization using a polarizer and the resulting beat signal between the two fields is detected using a fast photodiode. The amplitude of the detected beat signal can be adjusted by a $\lambda/4$ mounted wave plate in front of the polarizer.

\begin{figure}[t!]

\includegraphics{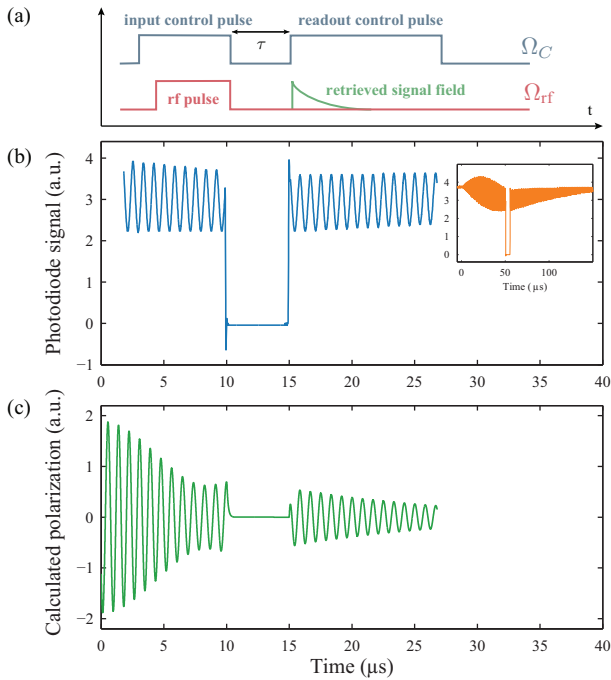}
\caption{\label{fig:rez}(a) Experimentally used pulse sequence for the resonant retrieval of stored coherence. (b) Photodiode beat signal between the optical control field and the coherent generated signal field for a 50 $\mu$s long rf pulse and the storage period of 5 $\mu$s. The inset shows a photodiode beat signal for a over a longer time span. (c) Calculated medium polarization on the transition between levels  ${|F = 1, m_F = 0\rangle\rightarrow |F' = 1, m_F = 1\rangle}$ (see Fig.~\ref{fig:6sys}), where the eigenfrequency was measured with the respect to the control field frequency. Parameters used in the calculation correspond to the estimated experimental parameters ($\Omega_C=0.02\gamma$, $\Omega_{rf}=0.0005\gamma$,
$g_F\mu_BB \propto 0.033\gamma$, $G=0.0001\gamma$, where $\gamma \approx 2\pi\times 6$ MHz).}
\end{figure}

The control laser power was on a 5 mm beam diameter typically set to 200 $\mu$W. The frequency of the applied radiofrequency was tuned to the Zeeman splitting between adjacent Zeeman sublevels near 1.2 MHz at the used level of the magnetic bias field, and the Rabi frequency of the rf field was approximately $\Omega_\text{rf}/2\pi=5$ kHz. Our experimental cycle, see Fig.~\ref{fig:rez}(a), begins by initially activating the linearly polarized control laser beam, so that population is pumped into the $|F = 1, m_F = 0\rangle$ component of the electronic ground state. We then in addition activate the resonant radiofrequency pulse which creates an atomic ground state coherence, after which the optical control field and the radiofrequency field are simultaneously turned off. After a 5 $\mu$s long period with no external driving fields applied, only the control field is reactivated, which retrieves the stored coherence and causes emission of a signal beam pulse \cite{Oberst}. This beam is generated by the coherence oscillating at the $|g_1\rangle$ to $|e\rangle$ transition of the simplified level scheme of  Fig.~\ref{fig:3sys}, or $|F = 1, m_F = 0\rangle$ to $|F' = 1, m_F =\pm1\rangle$ of the full level scheme (Fig.~\ref{fig:6sys}), and upon reactivating the control beam emission of a second optical beam collinear to the control beam is possible. The blue line in Fig.~\ref{fig:rez}(b) shows typical experimental data for the observed beat signal between the control and the signal beam. Fig.~\ref{fig:rez}(b) shows that a beat signal, as understood from emission of a signal field, is not only present during the retrieval of the coherence, but also during the preparation stage, where both the radiofrequency field and the control beam are active. As in that phase we deal with a driven system, we expect that before the storage the beat frequency will be determined by the frequency of the rf field, which in the presence of a finite detuning differs from the atomic eigenfrequency. The frequency of the detected beat, both during the rf pulse and after the coherent retrieval, was determined by fitting the signal with a sinusoidal function.

To allow for a comparison with expected results, we have carried out numerical simulations of our system using time-dependent density matrix calculations. In the simulations, a 6 level system, corresponding to the ground and excited states of the experimentally used $F=1 \rightarrow F'=1$ transition, was considered. The time evolution of the density matrix $\rho$ is calculated by using the Liouville equation: 
\begin{equation}
\frac{\partial\rho}{\partial t}=-\frac{i}{\hbar}\left[H, \rho\right]-\Gamma(G,\gamma)\rho
\end{equation}
where H is the Hamiltonian of the system that includes interactions with the optical control field, the external DC magnetic field and the rf field. The superoperator $\Gamma$ incorporates both a dephasing rate $G$ and population relaxation rate $\gamma$ in a phenomenological way. For the decoherence rate $G$ of our Rb buffer gas system, in the calculations we used the experimental value derived from EIT resonance measurements, as in \cite{Figueroa}. To mimic the experimentally existing population losses from the excited hyperfine level $5 P_{1/2}\ F'=1$ to the level $5 S_{1/2}\ F=2$, one additional element of the density matrix $\rho$ is introduced phenomenologically. As usual, the diagonal elements of the density matrix give the population of the states, while off-diagonal elements can be used to calculate the complex polarization of the medium \cite{Fleisch2005}. A typical result is shown in Fig.~\ref{fig:rez}(c), for parameters corresponding to the experimental settings.

\begin{figure}[t!]
\centering
\subfigure[\ ]{
\includegraphics[scale=1]{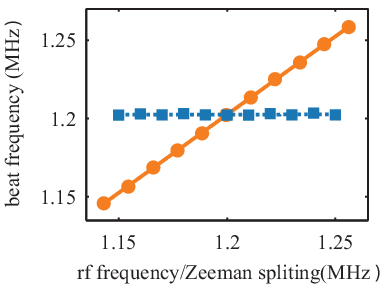}\label{fig:expscan}}
\subfigure[\ ]{
\includegraphics[scale=1]{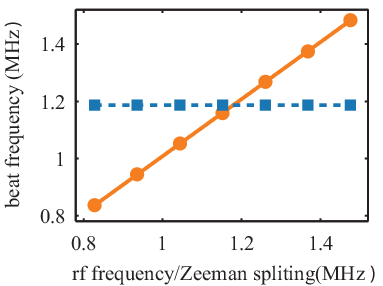}\label{fig:simscan}}
\subfigure[\ ]{
\includegraphics[scale=1]{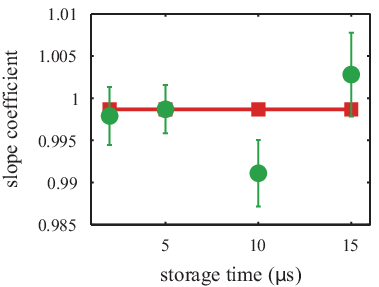}\label{fig:storage}}
\subfigure[\ ]{
\includegraphics[scale=1]{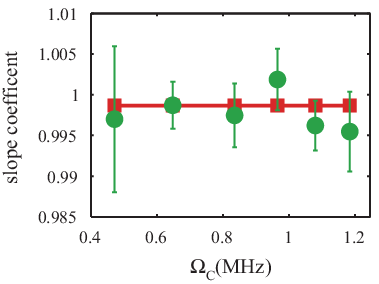}\label{fig:opt}}
\caption{(a) Measured beat frequency versus frequency of the radiofrequency field (blue squares) and Zeeman frequency splitting between adjacent Zeeman sublevels (orange dots) When the magnetic bias field was varied the frequency of the rf field was kept constant and vice versa. Each shown data point represents the averaged value of ten measurements. (b) Corresponding calculated results from the numerical simulation. (c) Slope of the observed beat frequency upon variation of the Zeeman splitting $g_F\mu_BB$ (green dots) versus the storage time, and corresponding simulated results for the atomic coherence (red squares). (d) As in (c), but instead of the storage time here the Rabi frequency of the optical control field $\Omega_C$ was varied. The size of the error bars of the simulated data values is within the drawing size of the data points.}
\end{figure}

In our experiments, we have detected the beat signal of control and generated optical signal beam for different values of both the magnetic bias field and the frequency $\omega_\text{rf}$ of the rf coupling. When the magnetic bias field was varied, the frequency of the rf field was kept constant and vice versa. Typical data for the shift in the beat frequency after the resonance retrieval, as a function of the bias field and the frequency $\omega_\text{rf}$ of the rf field is shown by orange dots and blue squares in Fig.~\ref{fig:expscan}. 
Here, the variable magnetic bias field is expressed in terms of Zeeman splitting of two neighboring ground states. Both parameters were scanned for the same frequency range, and each shown data point represents the averaged value for ten different measurements. These two data sets are fitted by a linear function, showing the linear dependence of the beat frequency on the size of the Zeeman splitting (Fig.~\ref{fig:expscan} solid orange line) with a slope of $0.999(\pm0.003)$. Within our experimental accuracy the beat frequency between optical control and generated signal fields thus equals the separation $g_F\mu_BB$ of adjacent Zeeman components of the $F=1$ hyperfine ground state. The data points corresponding to different values of the frequency $\omega_\text{rf}$ of the rf field can well be described by linear behavior (Fig.~\ref{fig:expscan} blue dashed line) with a slope of $0.002(\pm0.005)$ that vanishes within the experimental uncertainty, which is consistent with the assumption that the frequency of the rf coupling not having a noticeable influence on the value of the retrieved beat frequency. As expected, during the rf excitation, the difference of the detected beat frequency is given by $\omega_S-\omega_C =\omega_\text{rf}$ when varying either magnetic bias field or the rf frequency.

Fig.~\ref{fig:rez}(c) shows calculated results for the coherences between the states $|F = 1, m_F = 0\rangle\rightarrow |F' = 1, m_F = 1\rangle$, where the eigenfrequency was measured with respect to the frequency of the optical control field. The corresponding polarization is responsible for the emission of the signal beam, and the shown oscillation is to serve as a numerical estimation of the expected time-dependence of the experimentally observed beat between signal and control optical fields respectively. Fig~\ref{fig:simscan} shows the corresponding calculated values of the oscillation frequency of the atomic polarization with respect to the control field frequency, versus the frequency of the rf field (blue squares) and the frequency splitting between adjacent Zeeman sublevels (orange dots). For determination of the corresponding frequency from the  numerical values for the polarization, a Fast Fourier transform procedure, implemented in a MATLAB environment, was used. The calculated results are in good agreement with the corresponding experimental observation (Fig.~\ref{fig:rez}(b) and Fig.~\ref{fig:expscan}). The experimentally measurable beat frequency range is several times smaller than the frequency range acquired by simulations, which is attributed to the experimentally limited detection sensitivity. 
 
We have also investigated the resonance retrieval effect for different storage times of the atomic coherence. The green dots in Fig.~\ref{fig:storage} show the measured slope of the observed beat signal upon variation of the Zeeman splitting between adjacent sublevels versus the storage time, and within experimental uncertainties no deviation from the expected value unity is observed. In red, results of the numerically calculated dependence is shown. Finally, we tested for the robustness of the effects by performing measurements for different intensities of the optical control field. Fig.~\ref{fig:opt} shows the observed (green dots) slope of the beat frequency upon variation of the Zeeman splitting versus the control field Rabi frequency, and in red results of a corresponding calculation are shown. For the used parameter range, within experimental uncertainties no frequency shift from the ac Stark shift was observed. Similarly, within experimental uncertain no dependence on the retrieved frequency difference was observed when varying the Rabi frequency $\Omega_\text{rf}$ of the rf field up to approximately a level of  $\Omega_\text{rf}/2\pi=100$ kHz.   

To conclude, we have investigated the storage of atomic coherence in a thermal vapor of rubidium atoms by an optical control field, and a radiofrequency field, followed by its coherent resonant retrieval. We observe that independently of the used frequency for storage, the difference frequency between the retrieved optical signal field and the control field matches the atomic ground state difference frequency. This is attributed to the atomic coherence in the absence of driving oscillating at an atomic eigenfrequency, i.e. that only phase (and amplitude), but no frequency information is stored. The here investigated optical-rf excitation scheme represents an experimental simplification with respect to schemes based on driving the atomic ensemble with different optical driving fields. Both our experimental and theoretical results demonstrate that the retrieved difference frequency in good accuracy is solely determined by the Zeeman splitting of atomic ground state sublevels. It is to be assumed that our experimental procedure can be used in measurements of the atomic transition frequencies without precise knowledge of the driving frequencies. Our results hold prospects for the development of simple and robust atomic clocks. Prospective applications in atomic magnetometry are also anticipated.

We acknowledge funding by the Deutsche Forschungsgemeinschaft (We 1748-11).


\bibliography{Mainfile}

\end{document}